\documentclass[epsfig,twocolumn,prl,aps,floats]{revtex4}
\usepackage{amsmath,amssymb}
\usepackage{graphicx,bm}
\usepackage{epsfig}

\begin{document}
\title{Mesoscopic competition of superconductivity and ferromagnetism: conductance peak statistics in metallic grains}
\author{S.\ Schmidt and Y.\ Alhassid}
\affiliation{Center for Theoretical Physics, Sloane Physics
  Laboratory, Yale University,  New Haven, Connecticut 06520, USA}
\begin{abstract}
We investigate the competition between superconductivity and ferromagnetism in chaotic ultra-small metallic grains in a regime where both phases can coexist. We use an effective Hamiltonian that combines a BCS-like pairing term and
a ferromagnetic Stoner-like spin exchange term. We study the
transport properties of the grain in the Coulomb blockade regime and identify signatures of the coexistence between pairing and exchange correlations in the mesoscopic fluctuations of the conductance peak spacings and peak heights.
\end{abstract}
\maketitle
Although superconductivity and ferromagnetism compete with each other, the coexistence of both states has recently been reported in heavy-fermion systems~\cite{heavyferm} and high-$T_c$ superconductors~\cite{highTc}. This observed coexistence has led to enormous renewed interest in those materials. A main objective is to find systems in which ferromagnetism and superconductivity can be easily transformed into each other by changing an experimentally controllable parameter.

In nano-sized metallic grains a coexistence regime has been predicted theoretically~\cite{ying,falci,schmidt}. It was shown that the presence of spin jumps in the ground-state phase diagram of a single grain is a unique signature of this coexistence and that an external Zeeman field can be used to control and tune the size of this regime~\cite{schmidt}.

Experimentally, the competition between superconductivity and ferromagnetism in  metallic grains has not yet been investigated, in part, because fabrication and control of nanoscopic metallic devices is very challenging. The first nanoscopic single-electron transistors (SETs) with superconducting islands were produced by breaking nanowires into small pieces~\cite{RBT97}. Discrete energy levels and pairing effects were observed in a single grain by measuring the tunneling conductance (for a review see~\cite{delft01}). Itinerant ferromagnetism has been studied as well in normal metallic grains such as Cobalt~\cite{ferro}. During the last decade, production and gating of metallic grains has been considerably improved, e.g., using chemically synthesized grains and electromigration~\cite{park,bolotin1}. A very recent development is the observation of superconductivity in doped silicon, which might further simplify the production and control of mesoscopic superconducting devices~\cite{busta}.

Here we calculate the tunneling conductance in the Coulomb blockade regime for an ensemble of almost-isolated metallic grains using a rate equation approach~\cite{al04}. Our analysis is based on an effective Hamiltonian for chaotic or disordered systems, which combines a superconducting BCS-like term and a ferromagnetic Stoner-like term originating in pairing and spin exchange correlations, respectively~\cite{univH,ABG02}. By renormalizing the pairing coupling constant, we can derive accurate results using exact diagonalization of the many-body Hamiltonian within a truncated band.

We propose that signatures of the coexistence of pairing and ferromagnetic correlations can be observed in the mesoscopic fluctuations of the conductance peaks in superconducting grains with a sufficiently strong exchange interaction. We find a regime in which the peak spacing distribution exhibits bimodality induced by pairing correlations, while the mesoscopic fluctuations of the peak heights are suppressed by the presence of ferromagnetic correlations. Comparison of experimental data and theoretical predictions for such systems might serve as an ideal testing ground for the interplay between the charge, spin and pairing channels in disordered systems.

\noindent {\bf Model.} We consider a metallic grain whose single-particle dynamics are chaotic and with a large dimensionless Thouless conductance $g_T$. In the absence of an orbital magnetic field, such a grain is described by an effective universal Hamiltonian of the form~\cite{univH,ABG02}
\begin{eqnarray}
\label{origH}
\hat H=\sum_{k\sigma}\hspace{-0.0cm}\epsilon_{k} c_{k\sigma}^\dagger c_{k\sigma}
+ \frac{e^2}{2C}\hat N^2 - G \hat P^\dagger \hat P  -  J_s \hat{\bf S}^2\;.
\end{eqnarray}
The one-body term in (\ref{origH}) describes the kinetic energy plus confining single-particle potential. Here $c_{k\sigma}^\dagger$ is the creation operator for an electron with spin $\sigma=\pm$ in the single-particle level $\epsilon_k$ lying within a band of width $g_T$.
 The second term on the r.h.s.~of (\ref{origH}) is a charging energy term which is constant for a grain with a fixed number of electrons $N$ ($C$ is the capacitance of the grain). The third term is the pairing interaction with strength $G$ and where $\hat P^\dagger=\sum_i c_{i+}^\dagger c_{i-}^\dagger$ is the pair creation operator. The fourth term in (\ref{origH}) is an exchange interaction expressed in terms of the total spin operator $\hat{\bf S}=\sum_{k\sigma\sigma'}c_{k\sigma}^\dagger {\boldsymbol \tau}_{\sigma\sigma'}c_{k\sigma'}/2$ ($\tau_i$ are Pauli matrices). The parameter $J_s$ is the exchange coupling constant (estimated values of $J_s$ for a variety of materials were tabulated in Ref.~\cite{gorok}).

The pairing interaction can only scatter time-reversed pairs from doubly-occupied to empty levels but does not affect the singly-occupied levels (referred to as ``blocked" levels). Thus each many-body eigenstate of the Hamiltonian (\ref{origH}) $\vert i\rangle=\vert {\cal U},\zeta ;{\cal B}, \gamma, S,M \rangle $ factorizes into two parts. One part $\vert {\cal U},\zeta\rangle$ is a zero-spin eigenstate of the reduced BCS Hamiltonian, $\sum_{k \sigma}\epsilon_k  c_{k\sigma}^\dagger c_{k\sigma} -G P^\dagger P$, constructed within a subset ${\cal U}$ of empty and doubly-occupied levels. Here $\zeta$ denotes a set of quantum numbers that distinguish the different eigenstates within ${\cal U}$.  For example, $\zeta$ can be defined as the quantum numbers of the Slater determinant that is obtained from $\vert {\cal U},\zeta\rangle$ in the non-interacting limit, i.e., $\lim_{G\to 0}\vert {\cal U},\zeta\rangle= \vert \zeta\rangle $. The other part of the eigenstate $\vert {\cal B}, \gamma, S,M
 \rangle $ is obtained by coupling the set of singly-occupied levels ${\cal B}$ (each of these levels has spin $1/2$) to total spin $S$ and spin projection $M$. Here $\gamma$ denotes a set of quantum numbers distinguishing between states with the same spin $S$~\cite{rupp,hakan}. For a specific set $\cal{B}$ of $b$ singly-occupied levels, the total spin ranges from $S=0$ ($S=1/2$) for an even (odd) number of electrons to $S=b/2$ with each spin value having a degeneracy of $d_b(S) = \binom{b}{S+ b/2} - \binom{b}{S+1+ b/2}$.

The only relevant energy scale of the reduced pairing Hamiltonian ($J_s=0$) is $\Delta/\delta$, where $\Delta$ is the bulk pairing gap and $\delta$ the single-particle mean-level spacing. The number of levels in the grain can be truncated by renormalizing $G$ such that the low-energy spectrum of the grain remains approximately the same. For a picketfence spectrum, the renormalized coupling constant is given by
$G_r/\delta=[{\rm arcsinh} \left((N_r+1/2)\delta/\Delta\right)]^{-1}$
where $N_r$ is the number of levels in the truncated band~\cite{al07}. Strictly speaking, this holds in the absence of an exchange interaction. However, since the exchange interaction affects only the blocked levels, the renormalization holds as long as the number of blocked levels is small compared with the total number of levels in the truncated band.

The pairing and exchange terms in (\ref{origH}) compete.  Pairing correlations tend to minimize the total spin of the grain while the exchange interaction favors maximal spin polarization. This competition leads to interesting and non-trivial behavior that is the focus of this work.

\noindent {\bf Conductance.} We consider a grain that is weakly coupled to leads. In the sequential tunneling regime a typical tunneling width $\Gamma$ satisfies $\Gamma \ll \delta, T $, and we assume the charging energy to be the largest energy scale $e^2/2C \gg J_s,\Delta,\delta,T$.  The conductance displays a series of sharp peaks as a function of gate voltage, each peak describing a tunneling event, in which the number of electrons in the dot changes from $N$ to $N+1$. In the rate equation approach, the conductance is determined solely by the many-body energies and tunneling rates $\Gamma_{ij}^{\rm l(r)}$ in the left (right) lead between an eigenstate $i$ of the $N$-electron grain and an eigenstate $j$ of the $(N+1)$-electron grain~\cite{al04}. These transition rates are given by
\begin{eqnarray}
\label{gdef}
\Gamma_{ij}^{\rm l (r)} = \Gamma_{0}^{\rm l(r)} |\langle \alpha' S' M'|
\psi_\sigma^\dag({\bf r}_{\rm l,r}) | \alpha S M \rangle |^2\;,
\end{eqnarray}
where we have used the notation $\alpha=({\cal U},\zeta, {\cal B}, {\bf \gamma})$ and introduced overall coupling strengths $\Gamma_0^{\rm l (r)}$.
The operator $\psi_\sigma^\dag({\rm \bf r}) = \sum_k \psi_k ({\rm \bf r}) c^\dag_{k \sigma}$
creates an electron with spin projection $\sigma$ at the left
(right) point contact ${\bf r}_{\rm l}$ (${\bf r}_{\rm r}$).
There is only one term in the sum over single-particle states $k$ that can contribute to the matrix element in (\ref{gdef}). This tunneling level (denoted by $\lambda$) is always a good quantum number, either in the final state $j=(\alpha',S',M')$ if tunneling occurs into an empty level (hence $n'_\lambda=1$) or in the initial state $i=(\alpha,S,M)$ if an electron tunnels into a singly-occupied level (hence $n_\lambda=1$). Since the total spin of the grain is conserved
we can apply the Wigner-Eckart theorem and express the matrix element in (\ref{gdef}) in terms of a reduced matrix element which is independent of the magnetic quantum numbers. In this case, we find for the reduced transition rates
\begin{eqnarray}
\label{redrates}
\tilde \Gamma_{\alpha S \alpha' S'}^{\rm l,r} = \Gamma_{\lambda}^{\rm l,r}\left\{ \begin{array}{cc} (2S'+1)\, |\langle {\cal U'},\zeta'|{\cal P}_\lambda\rangle|^2 & {\rm if}\;\;\; n_\lambda'=1 \\ (2S+1)\, |\langle {\cal P}'_\lambda |{\cal U},\zeta\rangle|^2 & {\rm if}\;\;\; n_\lambda=1  \end{array} \right.
\end{eqnarray}
with $\Gamma_{\lambda}^{\rm l,r}=\Gamma_{0}^{\rm l,r}\left|\psi_{\lambda}({\rm \bf r}_{\rm l,r})\right|^2$. The first case $n'_\lambda=1$ is characterized by ${\cal B}'={\cal B} +\{\lambda\}$ and ${\cal U}'={\cal U} -\{\lambda\}$.  The state $|{\cal P}_\lambda \rangle= \hat R_\lambda \hat P_{n_\lambda=0}|{\cal U},\zeta\rangle$ is obtained from the Slater determinant expansion of $|{\cal U},\zeta\rangle$ by keeping only those determinants in which the level $\lambda$ is empty (defined by the projector $\hat P_{n_\lambda=0}$) and removing this level (defined by the operation  $\hat R_\lambda$). In the second case $n_\lambda=1$ we have  ${\cal B}'={\cal B} - \{\lambda\}$, ${\cal U}'={\cal U} +\{\lambda\}$, and $|{\cal P}'_\lambda \rangle= \hat R_\lambda \hat P_{n_\lambda=2}|{\cal U}',\zeta'\rangle$. The rates in Eq.~(\ref{redrates}) are used as an input for a reduced system of rate equations, which is obtained from Eqs.~(9a,b) in Ref.~\cite{al04} by replacing the full indices $(i,j)$ with the reduced indices $(\alpha S,\alpha' S')$.

In a chaotic grain (without a time-reversal symmetry-breaking magnetic field), the single-particle levels and wave functions follow the Gaussian orthogonal ensemble of random matrix theory~\cite{al00}. We study the statistics of the Coulomb-blockade conductance peaks at a typical experimental temperature of $T=0.1\delta$~\cite{delft01}.
For each realization of the one-body Hamiltonian, we calculate the five lowest many-particle eigenstates of the many-body Hamiltonian (\ref{origH}) using the Lanczos method for a bandwidth of $N_r=8$ and up to $19$ electrons. Using the many-body energies and wave functions as input, we calculate the transition matrix elements (\ref{redrates}) and solve the system of reduced rate equations. For each tunneling event, the maximal value of the conductance as a function of gate voltage is found numerically. The above procedure is repeated for  $\sim 4000$ samples to collect sufficiently good statistics. In the limit $\Delta=0$ we compare with larger bandwidth results obtained from a closed solution of the rate equations~\cite{rupp} and find good agreement. We have also verified that the contribution of higher many-body eigenstates is negligible for $T\alt 0.1\delta$.

\begin{figure}[t]
\epsfxsize=0.75\columnwidth{\epsfbox{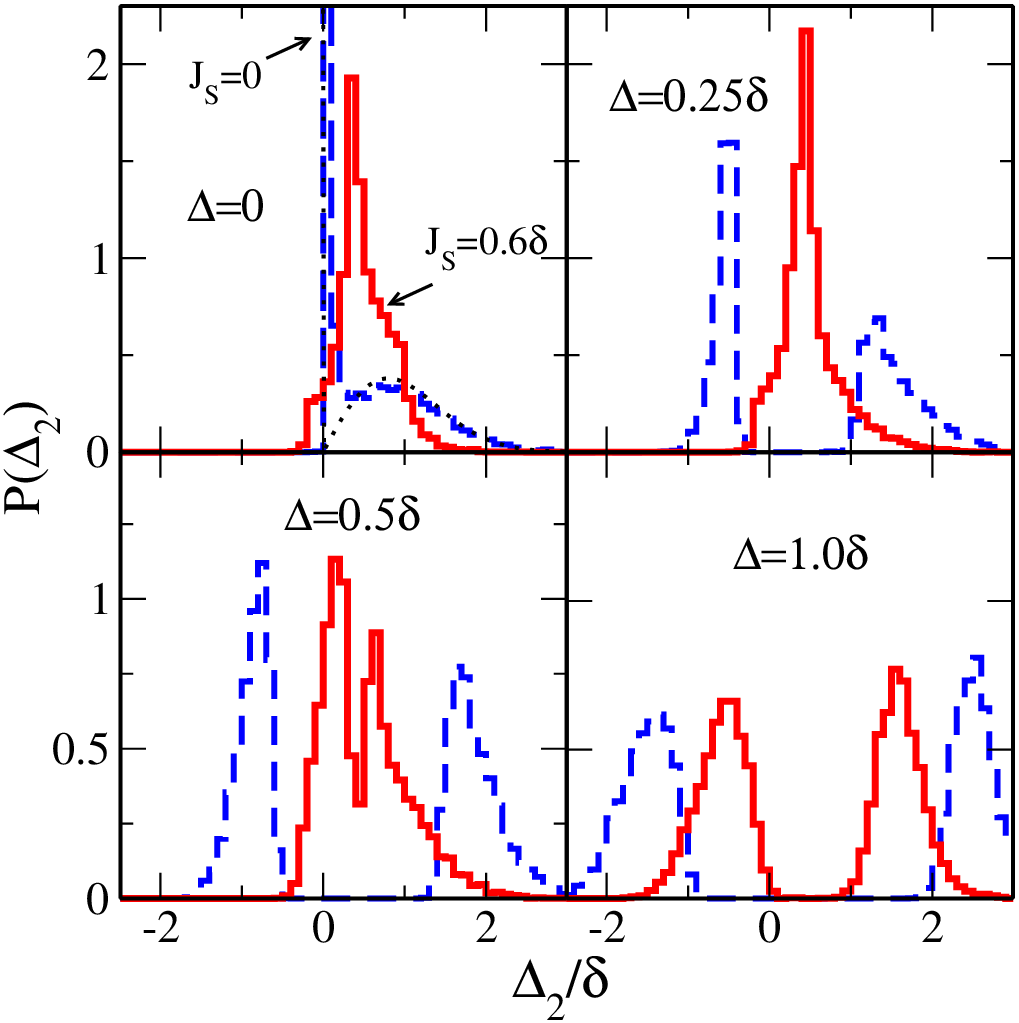}} \caption{\label{spacing} Peak spacing distributions at $T=0.1\delta$ for several values of the pairing ratio $\Delta/\delta=0, 0.25, 0.5, 1$. Each panel shows results for both $J_s=0$ (dashed histograms) and $J_s=0.6\delta$ (solid histograms). The dotted line in the top left panel is the analytic distribution for $\Delta=J_s=0$ and $T \ll \delta$. }
\end{figure}

\noindent{\bf Peak spacing statistics.} The peak spacing $\Delta_2$ measures the distance between two neighboring conductance peaks. Conductance peak-spacing distributions are shown in Fig.~\ref{spacing} for different values of $\Delta$ and $J_s$. In the absence of pairing and exchange interactions ($\Delta=J_s=0$), the distribution is bimodal because of the spin degeneracy of the single-particle levels. The exchange interaction tends to suppress this bimodality since it leads to spin polarization and thus mesoscopic fluctuations of the spin. At $J_s/\delta=0.6$, the bimodality is completely washed out.

The most significant effect of pairing correlations on the peak spacing distribution is to restore bimodality. For an exchange coupling of $J_s/\delta=0.6$, bimodality starts to reappear for $\Delta/\delta = 0.5$.  At even stronger pairing $\Delta/\delta=1$, the peak spacing distribution is well separated into two parts.  The left part describes the sequence of even-odd-even (EOE) tunneling events, while the right part corresponds to odd-even-odd (OEO) transitions. For $\Delta/\delta \geq 1$ the ground state spin of an even (odd) grain is almost always $S=0$ ($S=1/2$) and mesoscopic fluctuations of the spin can be ignored. Increasing the exchange coupling constant merely shifts the two parts of the distribution closer together.

The separation of the peak spacing distribution into two parts in the presence of pairing correlations can be understood qualitatively using the fixed-$S$ BCS  approximation at $T=0$ ~\cite{schmidt} for a picketfence spectrum.  For an EOE sequence the first peak corresponds to the blocking of an additional single-particle level and the second peak to the removal of this blocked level by creating an additional Cooper pair. An expansion of the BCS ground-state energy in powers of $\delta/\Delta$ yields to leading order
\begin{eqnarray}
\label{eoe_gap}
\Delta_2^{\rm (EOE)}=-2\Delta+\delta+(3/2)J_s+{\cal O}(\delta^2/\Delta)\,.
\end{eqnarray}
In an OEO tunneling sequence, one first creates a Cooper pair and then blocks a single particle level. In this case
\begin{eqnarray}
\label{oeo_gap}
\Delta_2^{\rm (OEO)}=2\Delta-(3/2)J_s+{\cal O}(\delta^2/\Delta)\,.
\end{eqnarray}

\begin{figure}[t!]
\epsfxsize=0.95\columnwidth{\epsfbox{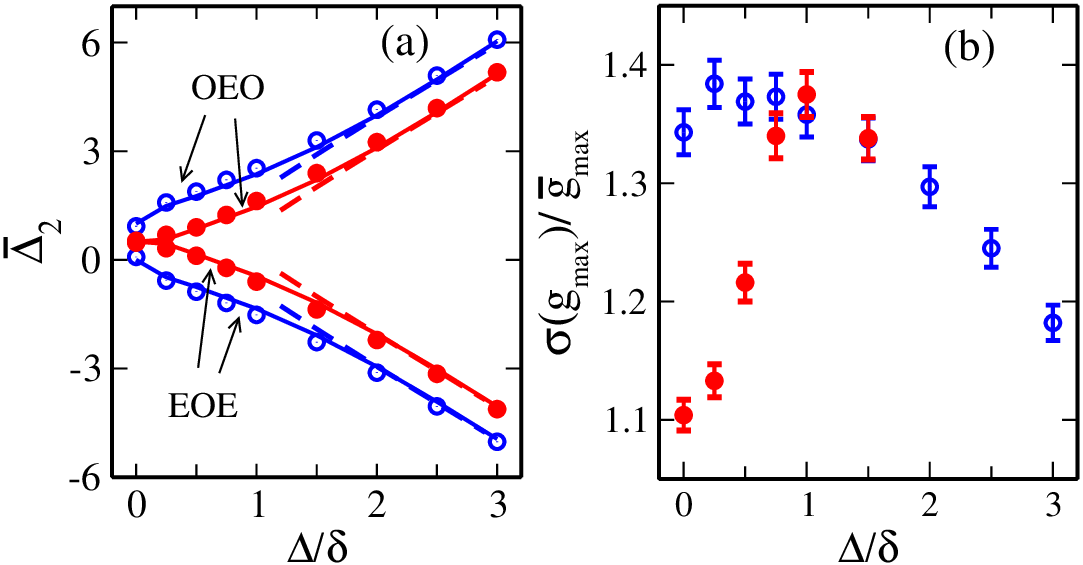}} \caption{\label{delta2} (a) Average peak spacing $\bar \Delta_2$,  and (b) relative peak height  fluctuation width $\sigma(g_{\rm max})/\bar g_{\rm max}$ as a function of the pairing ratio $\Delta/\delta$ for $J_s=0$ (open circles) and $J_s=0.6\delta$ (solid circles) at temperature of $T = 0.1 \delta$ . In panel (a) we compare the results with the value of $\Delta_2$ for a picketfence spectrum (solid lines) and the fixed-$S$ BCS mean-field theory (dashed lines). The error bars in panel (b) are the estimated statistical errors using 4000 random-matrix samples.
}
\end{figure}

In Fig.~\ref{delta2} we show that Eqs.~(\ref{eoe_gap}) and (\ref{oeo_gap}) are a good approximation to the exact average peak spacing $\bar \Delta_2$  for $\Delta/\delta>2$. Since the pairing gap enters in Eqs.~(\ref{eoe_gap}) and (\ref{oeo_gap}) with opposite signs, the separation of the two parts of the distribution becomes larger with increasing pairing gap. This effect is reduced in the presence of an exchange interaction as is also seen in Fig.~\ref{delta2}a. The exchange interaction strength $J_s/\delta$  is a material constant and for most metals is smaller or comparable to $J_s/\delta\sim 0.6$ ~\cite{gorok}. In contrast, the pairing ratio $\Delta/\delta$ can be made larger by using a smaller grain. Thus the bimodality effect can be tuned experimentally by measuring grains of different sizes. We also observe in Fig.~\ref{delta2} that a picketfence spectrum provides a good approximation for the mesoscopic average of $\Delta_2$. While the results presented here are for $T=0.1\delta$, the peak spacing statistics do not change significantly at lower temperatures.

\begin{figure}[t]
\epsfxsize=0.8\columnwidth{\epsfbox{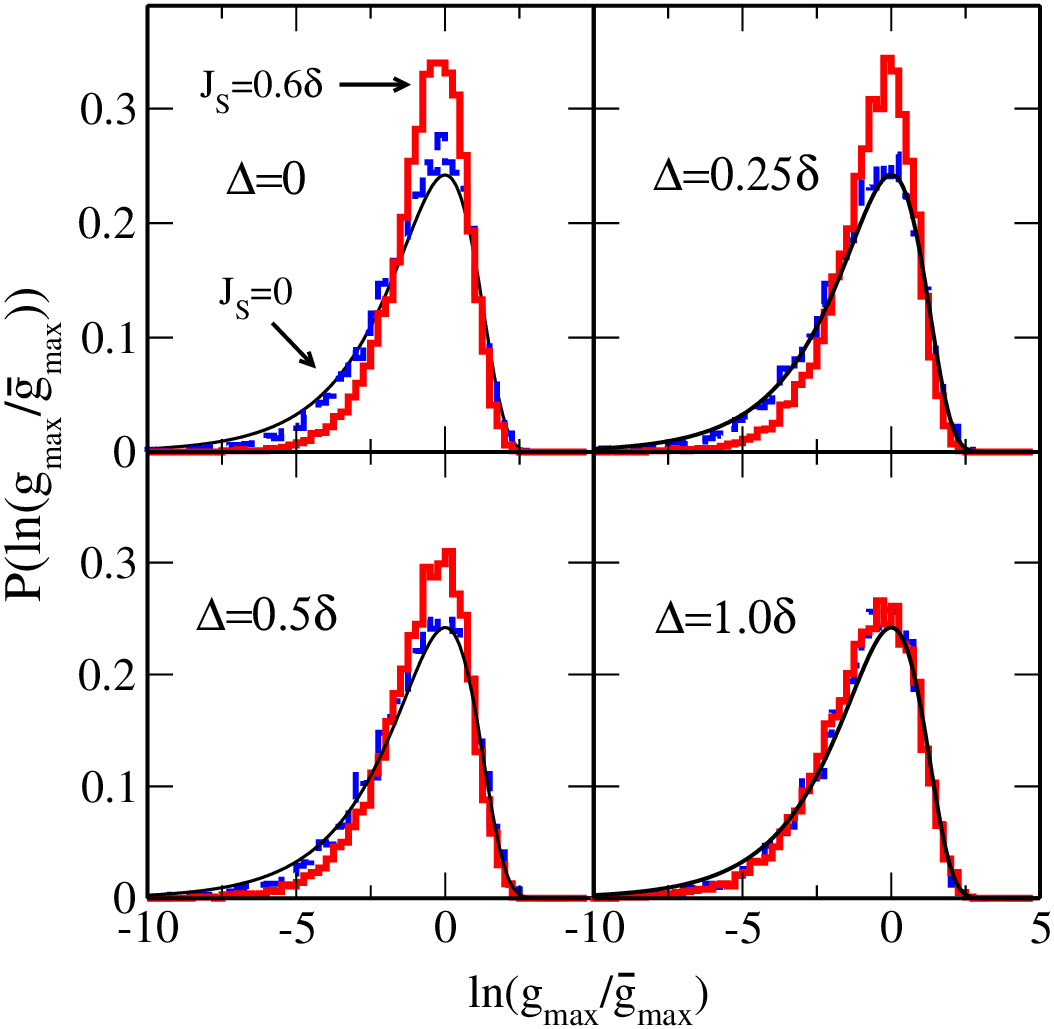}}
\caption{\label{peaks} Conductance peak height distributions $P(\ln(g_{\rm max}/\bar g_{\rm max}))$ at $T=0.1\delta$ in the presence of pairing and exchange correlations (for notation see Fig.~\ref{spacing}). We compare with the analytic distribution  for $\Delta=J_s=0$ and $T \ll \delta$~\cite{jalabert92} (solid lines).}
\vspace{-0.5cm}
\end{figure}

\noindent{\bf Peak height statistics.} Peak height distributions are shown in Fig.~\ref{peaks} [we use $\ln(g_{\rm max}/\bar g_{\rm max})$ where $\bar g_{\rm max}$ is the average conductance peak height]. In the absence of exchange and pairing interaction ($\Delta=J_s=0$) the distribution is known analytically at $T \ll \delta$ ~\cite{jalabert92}. The effect of exchange interaction on the width of the peak height fluctuations is somewhat similar to the effect of temperature. While temperature increases the number of levels contributing to the conductance, the exchange interaction brings down high-lying spin states and thus increases the number of states contributing to the conductance at small but {\em finite} temperatures.  Consequently, the ratio $\sigma(g_{\rm max})/\bar g_{\rm max}$ between the standard deviation and average of the conductance peak height decreases in the presence of exchange (see Fig.~\ref{delta2}b). In particular, we observe (see Fig.~\ref{peaks} for $\Delta=0$) that the exchange interaction suppresses the occurrence of small peak heights~\cite{very-low-T}. This suppression was observed in semiconductor quantum dots, where the Cooper pairing channel can be ignored and a closed solution for the conductance is available \cite{rupp}. There, the finite-temperature suppression of the probability of small conductance peak heights induced by exchange correlations led to a better agreement between theory and experiment.

Pairing correlations induce a gap that pushes larger spin states to higher energies. Thus, we observe that the ratio $\sigma(g_{\rm max})/\bar g_{\rm max}$ (Fig.~\ref{delta2}b) and the probability of small conductance peak heights (Fig.~\ref{peaks}) increase with $\Delta$ (up to $\Delta/\delta \sim 1$) for $J_s=0.6\delta$. For a pairing interaction that is strong enough to destroy all spin polarization (e.g., $\Delta/\delta \geq 1$ for $J_s=0.6\delta$), the peak height distribution becomes essentially independent of the exchange interaction strength  and the suppression effect of $\sigma(g_{\rm max})/\bar g_{\rm max}$ by exchange correlations completely disappears (see Fig.~\ref{delta2}b).

\noindent{\bf Mesoscopic coexistence.} A mesoscopic coexistence regime of superconductivity and ferromagnetism  can be defined by the simultaneous occurrence of a bimodality in the peak spacing distribution (a signature of pairing correlations) and the suppression of peak height fluctuations (a signature of ferromagnetic correlations). In Figs.~\ref{spacing} and \ref{peaks} such a coexistence case corresponds to  $\Delta/\delta=0.5$ and $J_s/\delta=0.6$. Interestingly, for a picketfence spectrum this case still belongs to the superconducting regime~\cite{schmidt}. Thus, we find that mesoscopic fluctuations increase the parameter regime for which coexistence of pairing and exchange correlations can be measured. The coexistence regime should be directly observable in Platinum, which has recently been shown to be superconducting in granular form~\cite{koenig} and also has a relatively large exchange interaction strength of $J_s/\delta \sim 0.59 - 0.72$~\cite{gorok}.
In experiments it might be difficult to vary the exchange interaction strength (a material constant). However, we expect to find similar effects at fixed exchange interaction strength by tuning an external Zeeman field.

\noindent{\bf Conclusion.} In this work we calculated the transport properties of an ultra-small metallic grain in its Coulomb blockade regime and studied the mesoscopic conductance fluctuations in the presence of both pairing and exchange correlations. Of particular importance is the finding that experimentally accessible mesoscopic signatures of superconductivity and ferromagnetism can coexist in a certain regime.

This work was supported in part by the U.S. DOE grant No. DE-FG-0291-ER-40608. Computational cycles were provided by the NERSC high performance computing facility at LBL and the Yale Bulldog cluster.
\vfill

\begin{thebibliography}{99}
\bibitem{very-low-T} This suppression effect disappears at very low temperatures ($T \alt 0.01 \delta$), where only the ground state contributes to the conductance and spin blockade can even enhance the occurrence of small peaks.
\end{thebibliography}
\end{document}